\documentclass[11pt,a4paper]{article}

\usepackage[T1]{fontenc}
\usepackage[utf8]{inputenc}
\usepackage{lmodern}
\usepackage[margin=1in]{geometry}
\usepackage{microtype}
\usepackage{amsmath,amssymb,amsthm,mathtools}
\usepackage[dvipsnames]{xcolor}
\usepackage{booktabs,tabularx,array,multirow}
\usepackage{enumitem}
\usepackage[hidelinks]{hyperref}
\usepackage{xurl}
\usepackage[capitalise,nameinlink]{cleveref}
\usepackage{caption}
\usepackage[section]{placeins}
\newcolumntype{L}[1]{>{\raggedright\arraybackslash}p{#1}}
\newcolumntype{Y}{>{\raggedright\arraybackslash}X}

\theoremstyle{definition}
\newtheorem{question}{Open Question}
\crefname{question}{Open Question}{Open Questions}

\newcommand{\F}{\mathbb{F}}

\newenvironment{digest}[1]{\par\smallskip\begingroup\raggedright\noindent\textbf{#1}\quad}{\par\endgroup}

\hypersetup{pdftitle={Structural Cryptanalysis of Classic McEliece},pdfauthor={Ramses Fernandez}}

\title{\textbf{Structural cryptanalysis of classic McEliece: Assessment of claims, costs and evidence}}
\author{Ramses Fernandez\textsuperscript{\href{https://orcid.org/0000-0002-8959-636X}{[0000-0002-8959-636X]}}\\[0.3em]
\small AegisQ Technologies\\
\small\texttt{rfernandez@aegisq.tech}}
\date{\today}

\begin{document}
\maketitle

\begin{abstract}
For the last two years, classical structural attacks on McEliece with binary Goppa codes progressed further, at the parameters of Classic McEliece, than in the previous four decades: a subexponential and then a quasi-polynomial distinguisher, key recoveries for public challenges with brute-force labels up to $254$, and conditional key-recovery estimates of about $2^{89}$ to $2^{128}$ per run of a large sparse linear-algebra computation, depending on the parameter set and on whether and how memory access is charged. On September 2026 the German BSI advised against new deployments of the scheme. This paper gives a version-aware critical assessment of these results. It separates proved distinguishers, conditional cost estimates and actual recoveries; analyses in detail Saarinen's HOVER algorithm, his conditional cost analysis and his recovery of the TII-254 challenge, together with Weis's cost study of Holdout/Hermite key recovery; and compares the other key-recovery routes and the objections raised against them. Our own analysis finds that the cost model is now the main quantitative uncertainty, with about thirty bits between the cheapest and the most expensive model for the same computation; that for the four larger parameter sets every model considered leaves the conditional estimates far below the targeted levels, whereas for \texttt{mceliece348864} the shortfall is of the order of the spread between models; that the structural estimates are lower than Grover-accelerated information-set decoding for the four larger sets in essentially every model and comparable for \texttt{mceliece348864}; and that none of this is a demonstrated attack, since the underlying heuristics have been tested only on instances tens of thousands to hundreds of thousands of times smaller than the target. We close with open questions.
\end{abstract}

\noindent\textbf{Keywords:} Classic McEliece; binary Goppa codes; structural cryptanalysis; key recovery; post-quantum cryptography.


\section{Introduction}\label{sec:intro}

\paragraph{Classic McEliece.} McEliece's cryptosystem of 1978, one of the oldest public-key cryptosystems, hides a binary Goppa code behind a disguised generator matrix \cite{McE78}; Niederreiter's dual version publishes a disguised parity-check matrix \cite{Nie86}. Its security has rested on two pillars. The first is the difficulty of decoding a random linear code: the best generic attacks, refinements of Prange's information-set decoding (ISD), have improved only slightly over six decades \cite{Prange62,BSI26}. The second is the difficulty of recognising or recovering the hidden Goppa structure: until recently the best key recovery for realistic parameters cost more than $2^{3000}$ operations \cite{Sen00,BSI26}. Classic McEliece, the modern Niederreiter-style key-encapsulation mechanism with binary Goppa codes, therefore sets its parameters by generic decoding \cite{ClassicMcEliece,ClassicMcElieceSec}. It was a fourth-round candidate in the NIST process that NIST did not select \cite{NISTIR8545}; selected parameter sets were standardised by ISO/IEC in June 2026 \cite{ISO18033Amd2,ClassicMcElieceISO}; and the German Federal Office for Information Security (BSI) has recommended the scheme since 2020 \cite{BSI26}.

\paragraph{The 2024--2026 wave.} Starting with Randriambololona's syzygy distinguisher in 2024 \cite{Ran25}, a sequence of classical structural attacks changed this picture within two years: higher-order-vanishing distinguishers and key recovery \cite{HW26,Hem26}, a polynomial-time attack at very high rates \cite{BLT26}, a heuristic subexponential key recovery \cite{BLRT26}, and the quasi-polynomial Holdout/Hermite distinguisher of Ghoshal, Ishai, Jain and Sun \cite{GIJS26}, on which a series of key-recovery algorithms and cost analyses has been built \cite{Ved26,Saa26cost,Wei26,Saa26b}. Public key-recovery challenges with brute-force labels up to $254$ were solved along the way \cite{BLT26,Hem26,Saa26a,Wei26,Saa26b}. The lowest published estimates for the Classic McEliece parameter sets are now conditional figures of about $2^{89}$ to $2^{128}$ per run of a large sparse linear-algebra computation, from bit operations with free memory to memory-charged costs, depending on the parameter set and the cost model \cite{Wei26}. On 29 September 2026 the BSI advised against new deployments of Classic McEliece, while recording that the published results do not yet enable a practical attack on its recommended parameter sets \cite{BSI26}.


\paragraph{Contributions.}
\begin{itemize}[leftmargin=*,itemsep=2pt]
  \item A structured account of the 2024--2026 structural attacks (\cref{sec:mceliece-wave} and \cref{tab:mceliece}) that separates proved distinguishers, conditional key-recovery estimates and recoveries of public challenges.
  \item Detailed analyses of Saarinen's three contributions, the HOVER algorithm, a conditional cost analysis of Holdout key recovery and the recovery of the TII-254 challenge (\cref{sec:mceliece-saarinen}), and of Weis's cost study, with its figures per parameter set and cost model (\cref{sec:mceliece-weis} and \cref{tab:weis}), re-checked against the cited versions.
  \item A comparison of the other key-recovery routes and of the objections raised against them, keeping version-specific claims apart (\cref{sec:mceliece-routes}).
  \item Our own assessment (\cref{sec:mceliece-verdict}): the cost model as the main quantitative uncertainty; the margins below the NIST levels for each parameter set; a comparison with Grover-accelerated ISD; a reconciliation of the figure quoted by the BSI with Weis's figures; the practicality of a full-size run; and what a repair would require. \Cref{sec:deployment} relates these findings to standardisation and deployment guidance, and \cref{sec:open} collects open questions.
\end{itemize}



\paragraph{Relation to a companion survey.} The attacks discussed here are classical; none uses a quantum algorithm or a hidden-shift structure. A companion survey \cite{FV26survey} asks which post-quantum assumptions would be affected by an efficient quantum algorithm for the dihedral coset problem. For binary Goppa codes it finds no attack certificate in the reductions it analyses, and it summarises the present paper for the purposes of its classification. The two papers can be read independently.

\paragraph{Organisation.} \Cref{sec:background} recalls linear codes, Goppa codes and Classic McEliece, generic decoding, the structural attacks before 2024, the cost models used later, and quantum considerations. \Cref{sec:mceliece-wave} surveys the 2024--2026 wave, and \cref{sec:mceliece-saarinen,sec:mceliece-weis,sec:mceliece-routes} analyse Saarinen's contributions, Weis's cost study and the other key-recovery routes. \Cref{sec:mceliece-verdict} gives our assessment, \cref{sec:deployment} turns to standardisation, deployment guidance and consequences, and \cref{sec:open} lists open questions and concludes.

\section{Background}\label{sec:background}

\subsection{Linear codes and decoding}\label{sec:bg-codes}

\paragraph{Decoding problems.} A linear $[n,k]$ code over $\F_q$ is a $k$-dimensional subspace $C\subseteq\F_q^n$, given by a generator matrix $G\in\F_q^{k\times n}$ or a parity-check matrix $H\in\F_q^{(n-k)\times n}$ with $C=\{c:Hc^{\top}=0\}$. The \emph{syndrome decoding} problem asks, given $H$, a syndrome $s$ and a weight $w$, for an error $e$ of Hamming weight $w$ with $He^{\top}=s$. Berlekamp, McEliece and van Tilborg proved its decision version NP-complete \cite{BMvT78}, whereas cryptographic security requires hardness of the particular average-case distribution. Worst-case NP-completeness alone does not establish average-case hardness for random codes, Goppa codes or deployed parameters.

The original McEliece scheme and Classic McEliece use algebraic codes, whose structure provides an efficient decoder. A generalised Reed--Solomon (GRS) code of dimension $k$ over $\F_q$, with distinct evaluation points $\alpha_1,\dots,\alpha_n$ and nonzero multipliers $y_1,\dots,y_n$, is $\{(y_1f(\alpha_1),\dots,y_nf(\alpha_n)) : f\in\F_q[x],\ \deg f<k\}$. It can be decoded efficiently up to half its minimum distance $n-k+1$, but its structure is easy to recover from any generator matrix \cite{SS92}; this is why McEliece-type schemes hide a code \emph{related} to a GRS code rather than a GRS code itself.

\subsection{Goppa codes, McEliece and Niederreiter}\label{sec:bg-goppa}

Let $q=2^m$, let $\alpha=(\alpha_1,\dots,\alpha_n)$ be distinct elements of $\F_q$, and let $g\in\F_q[x]$ be a polynomial of degree $t$ with $g(\alpha_i)\neq0$ for all $i$. The binary Goppa code $\mathrm{Goppa}(\alpha,g)$ consists of the vectors $c\in\F_2^n$ with
\begin{equation}\label{eq:goppa}
  \sum_{i=1}^{n}c_i\frac{\alpha_i^j}{g(\alpha_i)}=0
  \quad\text{in }\F_{q},\qquad 0\leq j<t.
\end{equation}
Each of these $t$ conditions over $\F_q$ amounts to $m$ binary conditions, so $\mathrm{Goppa}(\alpha,g)$ has dimension $k\geq n-mt$, with equality for the parameters considered here. The conditions \eqref{eq:goppa} say that $\mathrm{Goppa}(\alpha,g)$ is the subfield subcode $C\cap\F_2^n$ of the dual $C$ of the GRS code of dimension $t$ with evaluation points $\alpha$ and multipliers $g(\alpha_i)^{-1}$: binary Goppa codes are alternant codes, and the subfield restriction is what hides the GRS structure. For square-free $g$, a binary Goppa code has an efficient decoder that corrects $t$ errors, which is what McEliece's cryptosystem exploits \cite{McE78}; the support $\alpha$ and the polynomial $g$ form the secret key. For square-free $g$ one also has $\mathrm{Goppa}(\alpha,g)=\mathrm{Goppa}(\alpha,g^2)$, so the code lies in a GRS code of dimension $n-2t$; this sharper inclusion is the starting point of the Holdout/Hermite attacks (\cref{sec:mceliece-weis}).

\paragraph{McEliece and Niederreiter.} In McEliece's 1978 cryptosystem the secret is such a binary Goppa code; the public key is a disguised generator matrix $G_{\mathrm{pub}}=SGP$ with $S$ invertible and $P$ a permutation; a message $\mu$ is encrypted as $c=\mu G_{\mathrm{pub}}+e$ with $e$ of weight $t$, and decrypted by undoing $P$, decoding, and undoing $S$ \cite{McE78}. Niederreiter's dual version publishes a disguised parity-check matrix and encrypts by encoding the message into the error vector, the ciphertext being its syndrome \cite{Nie86}. Classic McEliece, a modern instantiation, is a Niederreiter-style KEM with binary Goppa codes and a public key in systematic form; its five parameter sets are listed in \cref{tab:params} \cite{ClassicMcEliece}. Classic McEliece was a fourth-round NIST candidate that NIST did not select \cite{NISTIR8545}; it was standardised by ISO/IEC in June 2026 and has been recommended by the BSI since 2020, a recommendation now restricted (\cref{sec:deployment}) \cite{ISO18033Amd2,BSI26}.

\begin{table}[htbp]
\centering\small
\renewcommand{\arraystretch}{1.12}
\begin{tabular}{@{}lrrrrcrr@{}}
\toprule
\textbf{Parameter set} & $m$ & $n$ & $t$ & $k=n-mt$ & \textbf{Category} & \textbf{pk (bytes)} & \textbf{ct (bytes)}\\
\midrule
\texttt{mceliece348864}  & 12 & 3488 & 64  & 2720 & 1 & 261\,120   & 96\\
\texttt{mceliece460896}  & 13 & 4608 & 96  & 3360 & 3 & 524\,160   & 156\\
\texttt{mceliece6688128} & 13 & 6688 & 128 & 5024 & 5 & 1\,044\,992 & 208\\
\texttt{mceliece6960119} & 13 & 6960 & 119 & 5413 & 5 & 1\,047\,319 & 194\\
\texttt{mceliece8192128} & 13 & 8192 & 128 & 6528 & 5 & 1\,357\,824 & 208\\
\bottomrule
\end{tabular}
\caption{The five Classic McEliece parameter sets of the fourth NIST round \cite{ClassicMcEliece}. The public key is the non-identity part of a systematic parity-check matrix, of $mt\lceil k/8\rceil$ bytes; a ciphertext has $\lceil mt/8\rceil$ bytes. Category: the NIST security category targeted \cite{ClassicMcEliece,Wei26}. The ISO/IEC amendment standardises selected sets and excludes \texttt{mceliece348864} \cite{ClassicMcElieceISO}.}
\label{tab:params}
\end{table}

\subsection{Generic decoding}\label{sec:bg-generic}

\paragraph{Information-set decoding.} The generic attack on syndrome decoding goes back to Prange \cite{Prange62}: choose $k$ coordinates at random, hope that the error vanishes on them, and solve the resulting linear system; each attempt succeeds with probability $\binom{n-k}{t}/\binom{n}{t}$, so the expected cost is exponential in $t$. Its refinements lower the exponent but keep the structure of a search over a large space of guesses. That structure is exactly what Grover's algorithm accelerates: Bernstein showed that a Grover search over Prange's guesses halves the exponent \cite{Ber10}, and Kachigar and Tillich obtained small further gains from quantum walks applied to the advanced variants \cite{KT17}. Until recently no structural attack came close, and the security levels of the scheme were set by generic decoding \cite{ClassicMcElieceSec}. \Cref{sec:mceliece-before,sec:mceliece-wave} describe how this changed.

For the five parameter sets, the estimates of \cite{GIJS26}, based on CryptAttackTester \cite{BC24}, put ISD at $2^{151}$ to $2^{287}$ bit operations with free memory. The Classic McEliece security guide uses another estimator: its smallest ISD figures are $2^{140.8}$ to $2^{275.6}$ without a memory charge, and $2^{158.6}$ to $2^{315.6}$ when the operation count is multiplied by the square root of the memory \cite{ClassicMcElieceSec,Wei26}. Given $2^{64}$ ciphertexts for one key, decoding one of them is cheaper by about $2^{32}$, the square root of their number \cite{Sen11,Wei26}. Grover's algorithm applied to Prange's ISD needs about $2^{101}$ to $2^{183}$ logical gates, ignoring error correction, memory and depth \cite{Ber10,Wei26}. \Cref{tab:weis} lists these figures for each parameter set.

\subsection{Distinguishers, key recovery and structural attacks before 2024}\label{sec:mceliece-before}

\paragraph{What constitutes a break?} Three kinds of structural result must be kept apart. A \emph{distinguisher} recognises that a public code belongs to a special distribution: it tells a McEliece public key from a uniformly random matrix of the same size. The textbook security argument for McEliece-type encryption combines the hardness of decoding random linear codes with the assumption that the public Goppa code cannot be distinguished from a random code \cite{FGOPT13}, and a distinguisher removes the second assumption from that argument. By itself, however, it supplies no decoder and does not violate the one-wayness assumption on which the security analysis of Classic McEliece rests \cite{BSI26,ClassicMcElieceSec}; a cryptosystem can survive a distinguisher. A \emph{key-recovery} attack must construct a decoding key from public information. It need not recover the original secret bit string: a different support and Goppa polynomial defining the same public code can give an equivalent decoder \cite{Saa26b}. For example, if the proposed support is $(\alpha_i)_{i=1}^n$ and the proposed polynomial is $g$, the binary vectors satisfying the conditions \eqref{eq:goppa} must define the public code, with distinct support elements, nonzero denominators and the required decoding conditions. This is a check on a recovered key. It does not prove that the procedure finding it succeeds efficiently on the full parameter distribution. Likewise, recovering one small challenge does not establish a cost bound for a standard parameter set. A \emph{message-recovery} attack, finally, decodes ciphertexts without the key; information-set decoding is the generic example. The 2024--2026 results are distinguishers and key-recovery attacks.

\paragraph{The situation before 2024.} Structural attacks on algebraic codes are as old as code-based cryptography. Sidelnikov and Shestakov recovered in polynomial time the support and multiplier of a generalised Reed--Solomon (GRS) code from a disguised generator matrix, breaking Niederreiter's original GRS-based proposal \cite{SS92}; their algorithm remains the final step of most modern attacks, which aim to expose a GRS code hidden inside the public data. A binary Goppa code is a subfield subcode of a GRS code, and the subfield restriction is what has protected it: it hides the algebraic structure that Sidelnikov--Shestakov exploits. For decades the best key-recovery attack on Classic McEliece parameters was therefore the brute-force search with Sendrier's support splitting algorithm, which decides whether two codes are permutation-equivalent \cite{Sen00}, at a cost that the BSI puts above $2^{3000}$ even for the smallest parameter set \cite{BSI26}. The first distinguisher, by Faugère, Gauthier-Umaña, Otmani, Perret and Tillich, worked only for codes of very high rate \cite{FGOPT13}, the regime of the CFS signature scheme of Courtois, Finiasz and Sendrier \cite{CFS01}. Couvreur, Mora and Tillich introduced a quadratic-form approach to distinguishing and attacking alternant and Goppa codes \cite{CMT23}; Bardet, Mora and Tillich gave a polynomial-time key recovery for high-rate random alternant codes \cite{BMT24}; and Kirshanova and May studied key recovery when part of the secret is known \cite{KM23}, a completion technique that the 2026 recoveries reuse. Until 2026 none of these applied to the Classic McEliece parameters at a cost anywhere near that of generic decoding \cite{Wei26}. In 2023 TII published a set of McEliece challenges to measure progress; their numerical labels are brute-force work-factor estimates, not security levels, a point that both of Saarinen's recovery papers stress \cite{TIIchal,Saa26a,Saa26b}.

\subsection{Cost models and security levels}\label{sec:bg-cost}

Concrete security claims compare the cost of an attack with that of a reference computation. NIST defines its security categories 1, 3 and 5 by key search on AES-128, AES-192 and AES-256; counted as gates, these searches give the reference levels $2^{143}$, $2^{207}$ and $2^{272}$ used in the cost analyses below \cite{NIST16,Wei26}. Three choices then separate published figures for one and the same attack. The first is the unit: bit operations executed by an algorithm, or gates of a Boolean circuit that does not depend on the key, the unit in which CryptAttackTester gives its ISD figures \cite{BC24}. The second is the treatment of memory: free; a charge of $M^\gamma$ for every operation of a computation that uses $M$ bits, with $\gamma=1/3$ or $1/2$, which is the rule behind the ISD tables of the Classic McEliece security guide \cite{ClassicMcElieceSec,Wei26}; or a charge per access, in which each operation pays $L^\gamma$ for the memory $L$ that it actually addresses, which for $\gamma=1/2$ is the cost on a two-dimensional mesh of small processors \cite{Ber01,LSTT02,Wei26}. The third is the algorithm assumed for the dominant step. The Holdout/Hermite attacks, which give the lowest estimates below, are dominated by kernel computations for very large sparse binary matrices, done by iterative block methods inherited from integer factoring: block Lanczos \cite{Mon95} and block Wiedemann \cite{Cop94}. The variant assumed by Ghoshal, Ishai, Jain and Sun, Eberly's block Lanczos algorithm for small fields, comes with a proof \cite{Ebe10}; Montgomery's algorithm over $\F_2$ does not \cite{Wei26}. \Cref{sec:mceliece-weis} shows how far these choices move the figures.

\subsection{Quantum considerations}\label{sec:bg-quantum}

Quantum algorithms enter in two ways. First, Grover's algorithm accelerates the search over guesses in ISD (\cref{sec:bg-generic}); NIST's categories account for such speed-ups through the comparison with key search on AES \cite{NIST16}. Second, we are not aware of a quantum structural attack competitive with the classical attacks discussed below; Weis knows of no quantum speed-up for the computation that dominates them \cite{Wei26}, and the hidden-subgroup approach to structure recovery meets proven limitations.

\paragraph{Quantum Fourier sampling.} In a code-equivalence formulation where both $G$ and $G_{\mathrm{pub}}$ are supplied, recovering a hiding pair $(S,P)$ can be expressed as a hidden-subgroup problem over a group built from $\mathrm{GL}_k(\F_2)$ and the symmetric group $S_n$. Dinh, Moore and Russell showed that, for well-permuted and well-scrambled codes, this hidden-subgroup problem cannot be solved by strong Fourier sampling, the measurement strategy behind the efficient abelian algorithms, or by any measurement of a single coset state, extending the negative results for graph isomorphism to subgroups of arbitrary structure, including automorphism groups of codes \cite{DMR11}, in line with the general limitations of coset states for $S_n$ \cite{MRS08,HMRRS10}. That formulation supplies the hidden generator $G$ as an additional input; it is not itself a key-recovery reduction from the McEliece public key alone. These results limit a class of measurements on coset states; they are not lower bounds against every quantum algorithm, and they do not by themselves establish quantum security.

Nor would an efficient quantum algorithm for the dihedral coset problem change this picture by itself: in one natural noise model, which does not cover adversarial bad registers, a noise-tolerant dihedral algorithm over $\F_2^n$ amounts to an algorithm for learning parity with noise, and the companion survey finds no attack certificate for binary Goppa codes in the dihedral reductions it analyses \cite{FV26survey}. All the attacks of \cref{sec:mceliece-wave,sec:mceliece-saarinen,sec:mceliece-weis,sec:mceliece-routes} are classical; whether a quantum computer accelerates their sparse linear algebra is open (\cref{sec:open}).

\section{The 2024--2026 wave}\label{sec:mceliece-wave}

\Cref{tab:mceliece} lists the results in the order in which they appeared, with the figures for Classic McEliece reported by their authors or by the BSI. Randriambololona's syzygy distinguisher \cite{Ran25} and the higher-order-vanishing (HOV) distinguisher of Hemmert and Wiemers \cite{HW26} exposed algebraic relations absent from random linear codes. Four lines of attack can be distinguished.
\begin{itemize}[leftmargin=*,itemsep=2pt]
  \item \emph{Syzygies.} The syzygy distinguisher detects excess higher syzygies in a shortened dual of the public code. According to the BSI it is the first algorithm to distinguish binary Goppa codes from general linear codes with a complexity asymptotically slightly subexponential in the error-correcting capacity $t$; its estimated cost for the parameter sets recommended in Technical Guideline TR-02102-1 is at least $2^{683}$ \cite{Ran25,BSI26}. Briaud, Lemoine, Randriambololona and Tillich turned a rank-two algebraic modelling into a heuristic key recovery that they conjecture to be subexponential; by its authors' lowest estimate, its distinguishing step alone still costs about $2^{454}$ at the level-1 Classic McEliece parameters \cite{BLRT26,Wei26}.
  \item \emph{Higher-order vanishing.} Hemmert and Wiemers, of the BSI, distinguish alternant and Goppa codes by the existence of homogeneous polynomials that vanish to higher order at the columns of a shortened public dual; for random codes of the same dimensions the corresponding kernel is zero \cite{HW26,Saa26a}. Hemmert turned the distinguisher into key recovery by searching, with a structured projective MinRank computation, for combinations of these polynomials whose first catalecticant has unusually small rank, combinations that omit one Frobenius block of the hidden GRS decomposition, and completing them to a Goppa key. He demonstrated the method on TII challenges and conjectured that it applies at Classic McEliece sizes at roughly the cost of the distinguisher, still above the targeted security levels \cite{Hem26,BSI26}. Saarinen's HOVER replaces the MinRank search by linear algebra (\cref{sec:mceliece-saarinen}) \cite{Saa26a}.
  \item \emph{Pfaffians and very high rates.} Bardet, Lemoine and Tillich gave a key recovery based on a Pfaffian modelling, polynomial in $m$ and $t$ for binary Goppa codes of very high rate \cite{BLT26,Saa26a}. It breaks the CFS signature scheme and recovered TII challenge keys up to TII-210, but it does not apply at the rates of Classic McEliece \cite{Saa26b,Wei26,BSI26}.
  \item \emph{Holdout/Hermite.} Ghoshal, Ishai, Jain and Sun gave a classical distinguisher that is quasi-polynomial in the asymptotic Classic McEliece regime and applies to every parameter set considered by NIST, with improved but not practical concrete estimates; a later version adds a heuristic key recovery \cite{GIJS26}. The lowest estimates all build on it: Vedenev's derivative-space attacks \cite{Ved26}, Saarinen's itemised costing \cite{Saa26cost}, Weis's sharpened distinguisher and two key recoveries \cite{Wei26}, and Saarinen's two-anchor recovery of TII-254 \cite{Saa26b}. Apon's lower bound concerns a step of an earlier Vedenev route \cite{Apo26}. \Cref{sec:mceliece-saarinen,sec:mceliece-weis,sec:mceliece-routes} analyse these works.
\end{itemize}
Asymptotic progress and successful attacks on high-rate codes are relevant evidence, but neither alone gives a practical attack on Classic McEliece.

\begin{table}[htbp]
\centering\footnotesize
\renewcommand{\arraystretch}{1.18}
\begin{tabularx}{\textwidth}{@{}L{2.8cm}L{1.8cm}Y L{3.85cm}@{}}
\toprule
\textbf{Work (posted)} & \textbf{Type} & \textbf{Idea and regime} & \textbf{Figure for Classic McEliece}\\
\midrule
Randriam\-bololona \cite{Ran25} (2024; EC 2025) & distinguisher & Excess syzygies of a shortened dual; subexponential in $t$ & $\geq2^{683}$, TR sets \cite{BSI26}\\
Hemmert--Wiemers \cite{HW26} (Sep 2025; CR 2026) & distinguisher & Higher-order vanishing (HOV) kernel nonzero for Goppa codes, zero for random codes & $\geq2^{550}$, TR sets \cite{BSI26}\\
Bardet--Lemoine--Tillich \cite{BLT26} (Mar 2026; CR 2026) & key recovery & Pfaffian modelling; polynomial time at very high rate; CFS, TII up to 210 & not applicable \cite{BSI26}\\
Briaud et al. \cite{BLRT26} (Jun 2026) & key recovery & Rank-two modelling; heuristic, conjectured subexponential & distinguishing step $\approx2^{454}$ at level 1 \cite{BLRT26}; $\geq2^{742}$ TR sets \cite{BSI26}\\
Hemmert \cite{Hem26} (Jun 2026) & key recovery & HOV directions found by projective MinRank search; TII challenges & conjectured $\approx$ distinguisher cost, above target \cite{Hem26,BSI26}\\
Ghoshal--Ishai--Jain--Sun \cite{GIJS26} (Aug 2026) & distinguisher (+ heuristic key recovery) & Holdout/Hermite; $n^{O(\log n)}$ asymptotically & $2^{114}$--$2^{124}$, five sets \cite{GIJS26}; TR sets $2^{120}$--$2^{124}$, key recovery $2^{140}$--$2^{148}$ \cite{BSI26}\\
Vedenev \cite{Ved26} (Aug 2026; rev.\ Oct 2026) & key recovery & Derivative spaces of the hidden curve; earlier routes run over $\F_4$; October revision: one-run finishing routes, TII-173, -249 and -253 & several distinguisher runs (earlier routes) \cite{BSI26}; $2^{89.9}$--$2^{98.6}$, conditional (October revision) \cite{Ved26}\\
Saarinen, HOVER \cite{Saa26a} (Aug 2026; rev.\ Oct 2026) & key recovery & Compatible-endomorphism algebra replaces MinRank search; five TII keys up to TII-252 & no threat in present form \cite{Saa26a}\\
Saarinen, holdout cost \cite{Saa26cost} (Aug 2026; rev.\ Sep 2026) & cost analysis & Itemised conditional cost of the GIJS locator route, one shortened instance & $2^{126.8}$--$2^{145.2}$, memory traffic excluded (\cref{tab:saarinen-cost})\\
Apon \cite{Apo26} (Aug 2026) & lower bound & A step of Vedenev's earlier jet-consistency route needs $>2t+3$ held positions & that route $>2^{1500}$ \cite{Apo26} for \texttt{mceliece8192128}\\
Weis \cite{Wei26} (Sep 2026; v0.5 Oct 2026) & distinguisher and key recovery & Sharper parameters; key recovery from gradients or from one run; cost model with memory; TII-253 & run $2^{89}$--$2^{98}$, or $2^{107}$--$2^{117}$ memory-charged (\cref{tab:weis}); $2^{99}$--$2^{102}$, TR sets \cite{BSI26}\\
Saarinen, TII-254 \cite{Saa26b} (Sep 2026) & key recovery & Two-anchor Holdout/Hermite; instance-specific & none claimed \cite{Saa26b}\\
\bottomrule
\end{tabularx}
\caption{Structural attacks on McEliece with binary Goppa codes since 2024, in order of appearance (EC: EUROCRYPT; CR: CRYPTO). Figures are bit-operation estimates as stated by the cited source; ``TR sets'' are the Classic McEliece parameter sets recommended by the BSI in TR-02102-1. Saarinen's and Weis's figures cover all five Classic McEliece parameter sets. The 2026 preprints have not completed peer review, and the BSI stresses that their estimates and cost models, for example for memory access, remain to be checked \cite{BSI26}; its figures predate some of the revisions listed. The lowest figures are conditional estimates per run of a large sparse linear-algebra computation, not end-to-end costs (\cref{sec:mceliece-weis}). For comparison, generic decoding costs $2^{151}$--$2^{287}$ bit operations for the five sets according to the estimates of \cite{GIJS26} based on CryptAttackTester \cite{BC24}, which Weis also uses \cite{Wei26}.}
\label{tab:mceliece}
\end{table}

\paragraph{The GIJS Holdout/Hermite method.} Ghoshal, Ishai, Jain and Sun \cite{GIJS26}, whose subtitle ``PIR meets McEliece'' refers to techniques from private information retrieval, exploit the fact that public columns arise from points on a hidden algebraic curve. They impose high-multiplicity vanishing conditions on low-degree polynomials at all but one column. For a Goppa key, the resulting relations also constrain the held-out column; a random matrix behaves differently. In the regime
\[
 m=\Theta(\log n),\qquad t=\Theta(n/\log n),\qquad k=\Theta(n),
\]
their distinguisher has time $n^{O(\log n)}$ and advantage $1-o(1)$. The quoted concrete distinguishing costs are $2^{114}$--$2^{124}$ bit operations over the five submitted parameter sets, $37$ to $163$ bits below generic decoding \cite{Wei26}. Their decryption and equivalent-key recovery extensions require additional heuristics; for the heuristic key recovery the BSI quotes $2^{140}$--$2^{148}$ at the TR-02102-1 sets \cite{BSI26}, while the costing in the revised GIJS paper, which already uses a sharper kernel count due to Weis, gives $2^{130}$--$2^{149}$ for the five sets \cite{Wei26}. Keeping this distinction matters: a theorem for distinguishing cannot silently certify the interpolation, rank and uniqueness assumptions of a recovery algorithm. \Cref{sec:mceliece-weis} describes the computation in detail.

\section{Saarinen's three contributions}\label{sec:mceliece-saarinen}

\paragraph{HOVER.} Saarinen's \emph{Higher-Order Vanishing Endomorphism Recovery} \cite{Saa26a} starts from Hemmert's key recovery and replaces its costly low-rank direction search by a linear-algebra computation. For a binary Goppa code with extension degree $m$, the scalar extension of the public dual decomposes into $m$ Frobenius-conjugate GRS blocks, and after shortening a suitable number of public coordinates the binary kernel $L$ of a degree-$p$ vanishing operator is nonzero for Goppa codes and zero for random codes of the same dimensions. Hemmert showed that $L\otimes\F_{2^m}$ contains determinantal forms $h_0,\dots,h_{m-1}$ whose projective lines form one Frobenius orbit, and found combinations omitting one GRS block through their small catalecticant rank, by a projective MinRank search that is estimated to test about $2^{m(e+2)}/m$ catalecticant ranks for an auxiliary index $e$. HOVER observes that the first catalecticants, the coefficient matrices of the first partial derivatives, of a basis of $L$ define a contraction tensor, and that the binary endomorphisms of $L$ compatible with every relation of this tensor form an associative algebra $\mathcal E_\beta$, computable by one binary linear system in $d^2$ unknowns, $d=\dim L$. Under the stated clean-case assumption, where $L\otimes\F_{2^m}$ is spanned by exactly the $m$ hidden directions and their images form a direct sum, $\mathcal E_\beta$ is a matrix algebra $M_c(\F_{2^{m/c}})$, with $c$ the common size of the classes of hidden directions sharing a kernel. When these kernels are pairwise distinct ($c=1$), $\mathcal E_\beta$ is isomorphic to $\F_{2^m}$ and the eigenspaces of any field generator are exactly the hidden directions, already ordered by Frobenius; for $c>1$ a residual projective search remains. The implementation is fail-closed: every unexpected algebra, failed field map or ambiguous phase returns no output, and only an exact row-space equality between the reconstructed and the public parity-check matrix accepts a key. Completion uses either the Segre--Veronese geometry of one recovered degree-three direction or the alignment of several shortenings, Schur products and Sidelnikov--Shestakov. In the revision of August 24, Saarinen reports public-key-only recoveries of TII-129, TII-213, TII-240, TII-246 and TII-252, with parameters $(n,m,t)$ up to $(1010,10,11)$: the three $t=11$ instances took between three and seven minutes on a laptop, and TII-129 and TII-213 between four and seven hours on a large-memory server; two further keys serve as controls. These are concrete structural recoveries, but the challenge label records an earlier brute-force estimate: ``TII-252 solved'' does not mean that $252$-bit security was broken. The paper is explicit about its limits: the generic termination of the clean algebra is unproved; the experiments establish seven exact instances, not a success probability over key generation; and for the five Classic McEliece parameter sets Hemmert's parameter rule gives vanishing degrees between $22$ and $52$ and a leading cost between about $2^{297}$ and $2^{690}$ operations for building the kernel that HOVER takes as input, a cost that HOVER does not change \cite{Hem26,Saa26a}. The revision of October 8 cited in the bibliography also reports five public TII key recoveries, including TII-252, and explicitly states that the present analysis does not indicate a threat to Classic McEliece parameters. HOVER should therefore receive credit for the algorithmic improvement and the experiments, without inheriting the much lower estimates of later Holdout-based methods.

\paragraph{An explicit conditional cost analysis.} Saarinen's separate costing paper \cite{Saa26cost} studies the GIJS direct locator-recovery route item by item. It uses one shortened coordinate set per parameter set, one shortened instance suffices because the completion step of Kirshanova and May needs only $mt+1$ support elements \cite{KM23,Wei26}, conditional on recovering sufficiently many mutually compatible locators, and it separates binary-coefficient matrix operations from general field arithmetic. Random row and column scaling and scalar Lanczos over a larger field provide quantified reliability for the linear solves; candidate solutions are mapped back and checked in the original equations. This solver reliability does not assign a probability to the structural assumptions. The analysis still requires suitable relation and block ranks, enough unique locator values consistent with one equivalent key, and a full-rank selection for key completion. Locally accepted choices need not be globally compatible. The paper also reduces the canonical relations of the GIJS key recovery to finite rank conditions, the reduction along which Weis later proves a conjecture of GIJS (\cref{sec:mceliece-weis}) \cite{Wei26}. \Cref{tab:saarinen-cost} records the estimates of the revision of September 15, the version that Weis compares with, rather than attributing Weis's lower figures to Saarinen.

\begin{table}[htbp]
\centering\small
\renewcommand{\arraystretch}{1.12}
\begin{tabular}{@{}lrrrr@{}}
\toprule
\textbf{Parameter set} & \textbf{Category} & \textbf{$\log_2 W$} & \textbf{$\log_2 S$} & \textbf{Reference exponent}\\
\midrule
\texttt{mceliece348864}  & 1 & 126.77 & 51.33 & 143\\
\texttt{mceliece460896}  & 3 & 145.22 & 59.10 & 207\\
\texttt{mceliece6688128} & 5 & 137.48 & 55.18 & 272\\
\texttt{mceliece6960119} & 5 & 136.65 & 54.86 & 272\\
\texttt{mceliece8192128} & 5 & 137.48 & 55.18 & 272\\
\bottomrule
\end{tabular}
\caption{Saarinen's conditional direct-locator estimates \cite{Saa26cost}, revision of September 15, 2026: $W$ is arithmetic work in bit operations; $S$ is working storage in \emph{bits}, not bytes. The last column gives the NIST classical-gate reference exponents used by that source. The comparison excludes address generation, memory traffic and communication, and includes unverified budget allowances. It is not a certified complete gate-count or elapsed-time attack.}
\label{tab:saarinen-cost}
\end{table}

The arithmetic estimates lie below the source's category references, but this is a conditional comparison in its chosen model. The stated recurrence, completion and storage allowances have not been validated by a full implementation at target size. The accompanying program recomputes dimensions and arithmetic budgets; it does not recover a Classic McEliece key. These figures are also not claimed to be optimal among recovery routes \cite{Saa26cost}.

\paragraph{The demonstrated TII-254 recovery.} The two-anchor Holdout/Hermite attack \cite{Saa26b} is a different result from both HOVER and the preceding cost table. It solves TII-254, the highest-labelled challenge solved at the time, with $(m,t,n)=(8,12,223)$: a binary $[223,127]$ code given by a full-rank $96\times223$ parity-check matrix, far smaller than \texttt{mceliece348864}. It builds on the Holdout/Hermite framework of \cite{GIJS26}, on Vedenev's derivative-space viewpoint \cite{Ved26} and on Hemmert's use of higher-order relations \cite{Hem26}, and it is stated as a thirteen-step process that separates heuristic choices, such as anchors and projections, from exact certificates. A degree-six relation operator, defined from Hasse-jet constraints in coordinates given by public columns, has about $2\cdot10^7$ source coordinates and $6\cdot10^9$ nonzero entries; its kernel contains the useful relations, hidden among a much larger space of universal identities. The attack proves and removes $4\,617\,031$ such nuisance directions by an exact coordinate-deflation theorem; conditions the remaining system on two public anchor columns; computes, by block Wiedemann \cite{Cop94} with width-$512$ Krylov sequences, a finite Hankel rank that proves the two conditioned kernels to have dimension exactly $121$; certifies their overlap ($106$), an $80$-dimensional pair core through a dual rank witness, and a common $64$-dimensional nuisance space; and finds that the quotient, of dimension $16$, behaves as a two-dimensional vector space over $\F_{256}$ whose one-dimensional subspaces are hidden projective locators. This yields $87$ of the $223$ support elements up to a projective change of coordinates; a bounded deterministic search and the completion method of Kirshanova and May \cite{KM23} recover the full support and the degree-$12$ Goppa polynomial, an equivalent Goppa key that an independent verifier checks against the public row space. The reported $27.2$ GPU-hours on NVIDIA GH200 accelerators cover the two final Krylov sequences, excluding reconstruction and CPU work; the whole effort, including the development of the method, took about eight days and fifteen hours of calendar time. These measurements are neither a total attack time nor a scaling law for Classic McEliece. The public artifact supports checking the recovered key and replaying the compact recovery stage, while its Lean proofs cover specified reusable linear-algebra steps. Again the paper is explicit about its limits: the process is not a general algorithm, even for other instances with the same parameters, since a different instance may fail to expose the $16$-dimensional carrier; TII-255, with $(m,t,n)=(8,24,249)$, remained unsolved; and Classic McEliece uses $m\in\{12,13\}$, $64\leq t\leq128$ and $3488\leq n\leq8192$, at which neither the degree-six construction nor the small residual kernels can be assumed to persist. The paper ``does not establish a practical attack on a Classic McEliece parameter set or an unconditional reduction of its claimed security level'' \cite{Saa26b}. It supplies concrete evidence for that instance, without proving full-size success or the conditional costs in \cref{tab:saarinen-cost}.

\paragraph{What Saarinen's papers establish.} Read together with the other results, the three papers support four conclusions. First, HOVER and the TII-254 paper are verified key recoveries for toy-sized codes, with $m\leq10$, $t\leq12$ and $n\leq1010$; the challenge labels they defeat are brute-force work factors, not security levels \cite{Saa26a,Saa26b}. Second, their contribution is methodological: both address what \cite{Saa26b} calls the common intermediate problem of recent attacks, converting publicly computable relations into a sufficiently precise description of the hidden code, and both replace heuristic acceptance by exact, replayable certificates (fail-closed algebra tests, finite-Hankel completeness bounds, Lean-checked lemmas, full row-space verification). This makes experimental cryptanalysis of structured codes more reliable, and it narrows the distance between distinguishers and key recovery. Third, the lowest estimates for Classic McEliece come neither from HOVER nor from the TII-254 recovery but from cost analyses of the Holdout/Hermite line: Saarinen's own conditional costing (\cref{tab:saarinen-cost}) and the lower figures of Weis and of Vedenev's October revision \cite{Saa26cost,Wei26,Ved26}. They concern an asymptotically quasi-polynomial distinguisher and key-recovery methods whose cost is that of a single run or of about $100$ to $1400$ runs of it, under assumptions that can only be tested on small keys. Fourth, the pace is itself a finding: estimates for structural key recovery on Classic McEliece parameters moved from more than $2^{3000}$ to conditional per-run figures near $2^{100}$ within a few months of 2026, after the syzygy distinguisher of 2024 had made distinguishing asymptotically subexponential in $t$ \cite{BSI26}. Saarinen credits the discovery of HOVER largely to large language models, the TII-254 implementation came out of a human-directed, AI-operated improvement cycle, and Weis's paper states that it was prepared with an AI system \cite{Saa26a,Saa26b,Wei26}. Whatever its long-term significance, this has shortened the cycle between an idea and a verified experiment; it also makes the version of a preprint part of any citation of its figures.

\section{The Holdout/Hermite computation, following Weis}\label{sec:mceliece-weis}

Weis's paper is the most detailed published account of what the Holdout/Hermite line costs for Classic McEliece, and the source of the figure that the BSI quotes; we summarise its version 0.5, posted on 6 October 2026, at some length \cite{Wei26}. One caveat governs every figure below. They are \emph{per-run} estimates for one sparse-linear-algebra computation: the two recovery routes use one run or roughly $100$ to $1400$ runs, and the figures retain four heuristic assumptions investigated on small keys, where the paper also discusses deviations. TII-253 is an actual reported recovery; full-size runs remain untested. A low per-run exponent must therefore not be reported as an unconditional end-to-end security level.

\paragraph{The computation.} All the attacks in \cite{Wei26} start from a fact shared by the recent literature. Let $q=2^m$, let $\alpha=(\alpha_1,\dots,\alpha_n)$ be distinct elements of $\F_q$ and let $\Gamma\in\F_q[Z]$ be square-free of degree $t$. The binary Goppa code $\mathrm{Goppa}(\alpha,\Gamma)$, of length $n$ and dimension $k=n-mt$, lies in a GRS code of dimension $D+1$, where $D=n-2t-1$: there are nonzero scalars $\lambda_j$ such that every codeword $c$ has a \emph{code polynomial} $f_c$ of degree at most $D$ with $c_j=\lambda_jf_c(\alpha_j)$ for all $j$ \cite{GIJS26}. If $y_1,\dots,y_n\in\F_2^k$ are the columns of a public generator matrix, this means that $y_j=\lambda_jp(\alpha_j)$ for a vector $p$ of $k$ polynomials of degree at most $D$: up to scalars, the public columns are points of a rational curve of degree $D$, written in unknown coordinates. The distinguisher of Ghoshal, Ishai, Jain and Sun first \emph{shortens} the code on $\ell$ coordinates, it keeps the codewords that vanish there and deletes these coordinates, and brings the generator matrix to systematic form. The resulting \emph{instance} is again a binary Goppa code, with $k_\ell=k-\ell$ unit columns and $mt$ non-unit columns, length $n_\ell=k_\ell+mt$ and curve degree $D_\ell=n_\ell-2t-1$. Fix a degree $d$ and consider homogeneous multilinear polynomials $h$ of degree $d$ in $k_\ell$ variables, that is, vectors of $N=\binom{k_\ell}{d}$ coefficients. One asks $h$ to vanish to order $s$ at the non-unit columns, meaning that all its Hasse derivatives of order below $s$ vanish there, where typically $s=d-1$; unit columns need no condition, because a multilinear form of degree $d$ vanishes to order $d-1$ at a unit vector. These linear conditions are the columns of a sparse binary \emph{jet matrix} $E$ with $N$ rows. One non-unit column $y_\tau$, the \emph{held-out column}, carries no condition. The test computes a nonzero $h$ in the row kernel of $E$ and checks whether $h(y_\tau)=0$. On a Goppa key the restriction $\varphi_h(Z)=h(p(Z))$ is a polynomial of degree at most $dD_\ell$, forced to vanish to order $s$ at the support elements of the constrained columns and to order $d-1$ at those of the unit columns; once these forced zeros outnumber the degree, $\varphi_h\equiv0$, and in particular $h(y_\tau)=\lambda_\tau^d\varphi_h(\alpha_\tau)=0$. On a random matrix a nonzero $h$ has $h(y_\tau)\neq0$ with probability at least $2^{-d}$. The computation of kernel vectors of $E$ by a block Krylov method is called a \emph{run}. Its cost grows quickly with $N$, so the aim is to take $d$ and $k_\ell$ as small as possible while keeping enough forced zeros and a nonzero kernel \cite[Sections 2--3]{Wei26}.

\paragraph{Sharper parameters.} Weis's first contribution lowers the cost of a run by about twenty bits and moves every parameter set from $(d,s)=(8,7)$ to $(7,6)$ \cite[Section 3]{Wei26}. Four observations contribute. The first is a count: in characteristic 2 the conditions of orders $d-2,d-4,d-8,\dots$ imply all the others \cite{HW26}, so a run keeps only these, and the average number of nonzero entries that one point contributes to a row of $E$ is then $7.4$ at $(7,6)$, instead of the bound of $120$ used by GIJS (about four bits). The second is a standard genericity heuristic of Fröberg type \cite{Fro85,HW26}, which predicts the dimension $\kappa$ of the row kernel; choosing parameters by $\kappa>0$, instead of GIJS's condition that $E$ have more rows than columns, saves about five bits, and that the row kernel has dimension at least $\kappa$ is Weis's Assumption 1. The last two are proved. For binary Goppa codes the products of $d$ code polynomials lie in an explicit space of dimension $dD_\ell+1-t$, which is $t$ less than for a generic GRS code (Proposition 1, which explains an observation of GIJS), so $t$ fewer forced zeros suffice: if $\Gamma$ is uniform among the monic irreducible polynomials of degree $t$, then, under mild conditions, a nonzero polynomial with the forced zeros exists with probability at most $n_\ell q^{-(t-\delta+1)}$, where $\delta\leq t$ is the shortfall of forced zeros against the degree (Theorem 3; below $2^{-39}$ at the parameters used). This saves eleven to twelve bits. And the sum $w$ of all columns of the instance is one more point of the curve, its point at infinity, at which conditions can also be imposed (Proposition 2). With these changes the distinguisher costs $2^{94.1}$--$2^{102.4}$ bit operations in the cost formulas of GIJS, against $2^{114.2}$--$2^{124.0}$ in their own analysis (\cref{tab:weis}).

\paragraph{Key recovery from gradients.} The distinguisher only uses the value of $h$ at the held-out column; its first derivatives there say more \cite[Section 4]{Wei26}. Since $h$ vanishes along the whole curve, its gradient at a point of the curve is orthogonal to the tangent there, and the tangent depends on the secret support. Theorem 4 makes this precise: if $\varphi_h\equiv0$ and only the value condition is imposed at $y_\tau$, then the vector of first-order derivatives $\nabla h(y_\tau)$, read as a message, encodes a codeword of the explicit subcode $C'_\tau=\mathrm{Goppa}(\alpha\setminus\{\alpha_\tau\},\Gamma\cdot(Z-\alpha_\tau))$ of the instance, which is defined by $m+1$ binary linear conditions. That the gradients of $k_\ell+16$ random kernel vectors span $C'_\tau$ is Assumption 2. Each run holds out $H$ columns, imposing only the value condition there, and $\lceil n_\ell/H\rceil$ runs, between $103$ and $1443$ for the five parameter sets, hold out every column once. Linear algebra in at most $n_\ell m\leq2.5\cdot10^4$ unknowns then matches the $m$ Frobenius-conjugate ``branches'' of the support across the columns, yields an explicit GRS code of dimension $3t+1$ for one branch, and reads the support as in Sidelnikov--Shestakov, up to a Möbius transformation $x\mapsto(ax+b)/(cx+e)$, which suffices for an equivalent key; the multipliers follow as in Wieschebrink's attack, and the shortened coordinates by the completion step of Kirshanova and May \cite{Wie10,KM23}. Starting from subcodes computed with the secret key, this finishing algebra rebuilt genuine \texttt{mceliece348864} and \texttt{mceliece8192128} keys in $40$ and $25$ minutes on one core.

\paragraph{Key recovery from a single run.} The gradient attack learns something only at the few columns that a run holds out. The single-run attack reads every constrained column at once from the column kernel of $E$, the space of linear relations among the conditions \cite[Section 5]{Wei26}. The distinguisher works because the value condition at a held-out column is a linear combination of the other conditions, on the curve, Hermite interpolation, and the coefficients of this interpolation relation depend on the nodes, hence on the support, much as in Sidelnikov--Shestakov. Ghoshal et al.\ proved this for GRS codes and conjectured it for binary Goppa codes; following a reduction by Saarinen \cite{Saa26cost}, Weis proves it (Theorem 5). If several held-out columns carry conditions of the same multiplicity $s'$, with $r=s'-1\geq2$ a power of two, and an exact ``no-slack'' count holds, then the column kernel contains two vectors $V_1,V_2$ over $\F_q$ with $V_2=\alpha_j^rV_1$ on a short ``locator block'' of every column $j$. Since $r$ is a power of two, $\alpha^r$ is a Frobenius conjugate of the support, so a single pair of vectors exposes the support element of every constrained column as a ratio. The obstacle is that almost all of the column kernel consists of \emph{local} relations, internal to one block of conditions or shared by two blocks, which carry no information about the code. GIJS remove them implicitly, by solving a fresh copy of the whole system for every column and every candidate value, about $n_\ell q$ large solves; Weis describes them in closed form (Proposition 6) and works modulo them, and that nothing else remains in the kernel is Assumption 3. The algorithm uses four held-out columns of multiplicity $s'=5$. One run returns $b=256$ random column-kernel vectors; after the local relations are removed column by column, three held-out columns normalised to $0,1,\infty$ fix the Möbius freedom, an $m\times m$ eigenvalue problem at the fourth selects one of the $m$ branches, and the support element of every other non-unit column is read as a ratio. One guessed support element, accepted or rejected by a polynomial greatest common divisor, then yields the Goppa polynomial and the unit columns. Weis estimates the post-processing at $2^{80}$--$2^{88}$ bit operations with a block Krylov method, about ten bits below the run; the estimate is untested, since his experiments used dense elimination.

\paragraph{What a run costs, and in which model.} The cost formulas of GIJS count bit operations for a solver chosen because it has a proof, Eberly's block Lanczos algorithm with two sparse preconditioners \cite{Ebe10}; by their storage formula a run needs $2^{57}$--$2^{62}$ bits, $96\%$ of them for the preconditioners. Weis's Section 6 organises the run differently and counts, for one and the same computation, its operations, its long-distance traffic and its memory \cite[Section 6]{Wei26}. The matrix $E$ is never stored: by Proposition 7 the product with $E$ decomposes into $2^{29}$--$2^{32}$ independent \emph{tiles}, each evaluating one quadratic form at the $P$ constraint points, so that a word arriving at a tile takes part in hundreds of additions and the long-distance traffic is about $2^8$ times smaller than the operation count. The solver is Montgomery's block Lanczos algorithm \cite{Mon95} with block width $b=256$, applied to $B=\tilde E^{\top}\Omega\tilde E$ with $\tilde E=E\Psi$, where $\Omega$ is a random involution of the rows and $\Psi$ a sparse invertible matrix with $\Psi\Psi^{\top}$ a random involution of the columns. $\Omega$ and $\Psi$ replace Eberly's preconditioners at negligible cost, but without a proof, and over $\F_2$ Montgomery's algorithm has none either. That the iteration terminates after about $\operatorname{rank}(E)/(b-0.76)$ steps and returns kernel vectors is Assumption 4; one instance at degree $7$ on which the plain variant, on $E^{\top}E$, failed was solved once both $\Omega$ and $\Psi$ were used. Weis then gives two counts of operations: $T$, the AND and XOR operations that the algorithm executes, which skips the zero entries of $E$ as GIJS's formulas do; and $T_{\mathrm f}$, the gates of a Boolean circuit that depends on the parameter set but not on the key, which is about three bits larger and is the count he compares with ISD and with the NIST levels, since the ISD figures also count the gates of a circuit \cite{BC24}. A run takes $T=2^{89.4}$--$2^{98.4}$ operations, or $T_{\mathrm f}=2^{92.3}$--$2^{101.3}$ gates, in $2^{30.1}$--$2^{34.1}$ dependent iterations on $2^{38}$--$2^{42}$ unknowns, with a memory of $M=2^{49.7}$--$2^{53.6}$ bits (\cref{tab:weis}). Memory is charged in two ways, an access to $M$ bits costing $M^{\gamma}$ with $\gamma=1/3$ or $1/2$. The rule of the Classic McEliece security guide \cite{ClassicMcElieceSec} multiplies the whole operation count by $M^{\gamma}$; it gives $T_{\mathrm f}M^{1/2}=2^{117.1}$--$2^{128.1}$ for the run, against $2^{158.6}$--$2^{315.6}$ for ISD under the same rule. A charge per access, $T_{\mathrm f}L^{\gamma}+V_{\mathrm f}M^{\gamma}$, charges each operation for the memory it actually addresses: $L^\gamma$ inside a tile of $L\approx2^{22}$ bits, and $M^\gamma$ for each of the $V_{\mathrm f}$ bits that travel between the tiles and the vectors. For $\gamma=1/2$ this charge is the cost on a two-dimensional mesh of small processors, whose area--time product grows like $N^{5/2}$, as for the matrix step of integer factoring \cite{Ber01,LSTT02}, and Weis regards it as the most meaningful single figure: $2^{107.3}$--$2^{117.2}$. No ISD figure with a charge per access exists, so he sets this best estimate against ISD with free memory, which favours ISD: the run is then $44$ bits cheaper for \texttt{mceliece348864} and $77$ to $170$ bits cheaper for the other sets, and it lies $36$ bits below the NIST level of \texttt{mceliece348864} and $93$ to $156$ bits below those of the others, the NIST levels being the gate counts $2^{143}$, $2^{207}$ and $2^{272}$ of the AES key searches that define categories 1, 3 and 5 \cite{NIST16}. The mesh also gives a time: at least $2^{58}$--$2^{64}$ sequential steps or, at $10^9$ steps per second with $b=256$, between $7$ years (\texttt{mceliece348864}) and $420$ years (\texttt{mceliece6688128} and \texttt{mceliece8192128}) on a machine with $100$ TiB to $1.5$ PiB of memory. In Weis's words, ``No machine of this kind exists''. More hardware saves little time, because the traffic does not depend on $b$: unlike exhaustive key search, or ISD with small lists, the run does not keep its cost when it is spread over more hardware.

\begin{table}[tb]
\centering\footnotesize
\renewcommand{\arraystretch}{1.12}
\begin{tabularx}{\textwidth}{@{}Y rrrrr@{}}
\toprule
& \multicolumn{5}{c}{\textbf{Classic McEliece parameter set (mceliece\dots)}}\\
\cmidrule(l){2-6}
\textbf{Quantity} ($\log_2$ of bit operations unless stated) & \textbf{348864} & \textbf{460896} & \textbf{6688128} & \textbf{6960119} & \textbf{8192128}\\
\midrule
NIST category & 1 & 3 & 5 & 5 & 5\\
NIST level (gates) \cite{NIST16} & 143 & 207 & 272 & 272 & 272\\
ISD, memory free \cite{GIJS26,BC24} & 151 & 191 & 257 & 257 & 287\\
\addlinespace
Distinguisher, GIJS \cite{GIJS26} & 114.2 & 120.3 & 124.0 & 123.0 & 124.0\\
Distinguisher, sharper parameters (b) & 94.1 & 99.3 & 102.4 & 101.6 & 102.4\\
Gradient attack (b): number of runs & 103 & 1443 & 236 & 294 & 236\\
Gradient attack (b): total & 100.8 & 109.8 & 110.2 & 109.8 & 110.2\\
Single-run attack (b) & 94.3 & 99.3 & 102.4 & 101.6 & 102.4\\
Single run, count $T$ (zeros skipped) & 89.4 & 95.4 & 98.4 & 97.6 & 98.4\\
Single run, fixed circuit $T_{\mathrm f}$ (gates) & 92.3 & 98.3 & 101.3 & 100.5 & 101.3\\
Memory $M$ ($\log_2$ bits) & 49.7 & 52.3 & 53.6 & 53.3 & 53.6\\
Iterations ($\log_2$) & 30.1 & 32.8 & 34.1 & 33.8 & 34.1\\
Run if Assumption 1 fails (count $T$) & 93.8 & 112.6 & 116.3 & 115.4 & 116.3\\
\addlinespace
Best estimate, $T_{\mathrm f}L^{1/2}+V_{\mathrm f}M^{1/2}$ & 107.3 & 114.0 & 117.2 & 116.4 & 117.2\\
Guide rule $T_{\mathrm f}M^{1/2}$: run & 117.1 & 124.5 & 128.1 & 127.2 & 128.1\\
Guide rule: ISD \cite{ClassicMcElieceSec} & 158.6 & 201.0 & 278.2 & 279.2 & 315.6\\
ISD for one of $2^{64}$ ciphertexts \cite{Sen11} & 119 & 159 & 225 & 225 & 255\\
Grover on Prange (logical gates) \cite{Ber10} & 101 & 124 & 164 & 164 & 183\\
Mesh time at $10^9$ steps/s (years) & 7 & 110 & 420 & 300 & 420\\
\bottomrule
\end{tabularx}
\caption{Cost figures of Weis \cite{Wei26}, version 0.5, for Holdout/Hermite distinguishing and key recovery on the five Classic McEliece parameter sets, with the generic attacks he compares them with (from Tables 1, 3, 4 and 5, Sections 6 and 9 and Appendix E of \cite{Wei26}). (b): the cost formulas of GIJS \cite{GIJS26} with Weis's parameters, which charge two operations per nonzero matrix entry and nothing for memory; $T$: Weis's count for his own organisation of the run (Montgomery's solver, products by tiles, slightly fewer unknowns), with one operation per nonzero entry, four to five bits below (b); $T_{\mathrm f}$: gates of a Boolean circuit that does not depend on the key; best estimate: each operation charged the square root of the memory it addresses, inside a tile of $L\approx2^{22}$ bits or, for the $V_{\mathrm f}$ bits of long-distance traffic, the whole memory $M$; guide rule: the whole operation count multiplied by $M^{1/2}$, as in the Classic McEliece security guide, whose smallest ISD entries are listed. The multi-target ISD and Grover figures carry no memory charge, and the Grover figures ignore error correction and depths of at least $2^{71}$ to $2^{150}$. The memory-charged figures are model estimates without experiment, and every figure rests on heuristic assumptions tested only on small keys.}
\label{tab:weis}
\end{table}

\paragraph{Assumptions, experiments and TII-253.} The numbered propositions and theorems of \cite{Wei26} are proved; the remaining ingredients are the four assumptions already met, on the dimension of the row kernel (Assumption 1), the span of the gradients (2), the structure of the column kernel (3) and the behaviour of the solver (4), which concern matrices with $2^{38}$ to $2^{42}$ unknowns and can only be tested on small keys; Table 7 of the paper lists what each figure rests on and how far each ingredient was tested. Weis's key-recovery experiments, in SageMath with the M4RI library and his own solver, cover degrees $5$ to $7$, degree $7$ being that of the attack on Classic McEliece (the kernel-dimension heuristic was also tested at degree $4$): both attacks end to end on toy keys ($m=6$, $t=6$, $n=64$); thirty keys in general position with $m=7$, $t=11$, $n=128$, all recovered, each from one run; ten runs at the threshold of the parameter conditions, all successful; and, at degree $7$, keys with $m=6$ and $t=4$, recovered from single runs when chosen in general position, whereas on a typical such key, with hundreds of short dependencies among its columns, the gradients still spanned the subcodes $C'_\tau$ but the single-run reading failed, as the paper's formula for such dependencies predicts. The deviations from the predicted dimensions had two sources. One is dual codewords of low weight, that is, short linear dependencies among the columns of a small shortened code; a formula of the paper predicts their effect, exactly for separate codewords and too high where many overlap, and for the shortened Classic McEliece codes their expected number is about $2^{-61}$ or less, while a search with Stern's algorithm \cite{Ste89} found none. The other is a few additional relations for small $t$, observed for $t\leq5$ and absent from $t=6$ on; Classic McEliece has $t\geq64$. Some ingredients were checked at full size without a run, among them the dimensions of the finishing algebra and the hypotheses of the single-run reading on generated keys of every shape. Untested are, among others, the behaviour of GIJS's solver assumed by the figures computed with their formulas, the cost of the post-processing, and every memory-charged figure, which is a model estimate without experiment. Weis also states what happens if an assumption fails. Assumption 1 is the only one that enters the choice of parameters; without it a run costs $2^{93.8}$--$2^{116.3}$ operations, four bits more for \texttt{mceliece348864} and seventeen to eighteen bits more for the others, still assuming Assumption 4. If Assumption 4 failed, the run would fall back on GIJS's solver, at the cost of the (b) figures of \cref{tab:weis} and without a memory-charged counterpart; and a systematic failure of Assumption 3 would defeat the single-run reading but not the gradient attack \cite[Section 9]{Wei26}. As a demonstration, Weis recovered the key of TII-253 ($m=8$, $t=9$, $n=214$), which had remained open while instances with labels up to $252$ had been solved \cite{Wei26,TIIchal}. At $(d,s)=(6,5)$ and $k_\ell=46$ the matrix has $9.4\cdot10^6$ rows, $1.18\cdot10^7$ columns and $3.1\cdot10^9$ nonzero entries, and the run took $35\,903$ iterations, or $21.7$ hours on $32$ threads of a shared server, about $2^{57}$ bit operations in the count of Section 6. All $68$ regular columns were read; the only deviation, a larger dimension in the post-processing at the seven non-unit columns of the instance's unique dual codeword of weight at most $10$, is exactly the one that the paper's formula predicts; a second run with other unit columns agreed with the first; and the organisers' verification script accepts the recovered key, which the paper lists in full.

\paragraph{Revisions of the estimates.} Because its figures are widely quoted, the revision history of \cite{Wei26}, given in its Appendix H, deserves a summary. Version 0.1 (11 September 2026) stated that no binary Goppa parameters of usable size resist the attack; the statement rested on example sets with $n=2^m$, which nothing requires, and was withdrawn in version 0.2 (18 September), which added the dependence of the cost on $m$. Version 0.3 (25 September) corrected a mistyped constant of the solver formulas, which raised some figures by $0.1$ to $0.3$ bits. Version 0.4 (5 October) proved two of the five assumptions of version 0.3 (now Theorems 3 and 5), restated the others, added the behaviour of the solver as Assumption 4, and replaced the earlier discussion of storage by the cost model of Section 6 and the comparison with ISD in several memory models. Version 0.5 (posted 6 October) changed no cost figure, but withdrew a statement of versions 0.1 to 0.3, that a run which does not store $E$ needs only $2^{47.8}$--$2^{52.1}$ bits without additional operations: it rested on an unpublished estimate for block Wiedemann \cite{Cop94}, never run, which charged one operation where the formulas charge two, and the memory-charged figures of those versions used both numbers. Memory-charged figures quoted from versions earlier than 0.4 should therefore be replaced by those of \cref{tab:weis}. The revision history records changes of at most $0.3$ bits to the figures computed with GIJS's formulas, and the paper thanks colleagues for identifying errors in earlier versions, among them errors in cost figures.

\paragraph{Dependence on the parameters.} The cost of a run depends on $m$ and $t$ but not on $n$, as long as $k\geq k_\ell$; at fixed $m$ it is polynomial in $t$. What raises it is $m$: the zero count forces the multiplicity $s$ to grow by about two units for every three units of $m$, and each unit of $s$ costs eleven to fourteen bits, so the cost grows by about eight bits per unit of $m$. The run is therefore quasi-polynomial in $n$ for families with $m=O(\log n)$, but not for families in which $m$ grows faster, and nothing in the McEliece design forces $n=2^m$ \cite[Section 9 and Appendix F]{Wei26}. Keeping the $t$ and the ISD cost of \texttt{mceliece348864}, \texttt{mceliece460896} and \texttt{mceliece6688128} while raising $m$, a run costs more bit operations than the corresponding NIST level from $m=19$, $27$ and $34$ on, with public keys of $0.65$, $2.2$ and $7.1$ MB, that is, $2.5$, $4.3$ and $6.8$ times the present sizes; with the square-root charge, $m=16$, $22$ and $29$ suffice, with keys of $0.47$, $1.5$ and $5.1$ MB. Weis presents these sets as illustrations of the scaling and explicitly not as proposals: they extrapolate Assumptions 1 and 4 far beyond the tests, and other attacks at large $m$ were not examined.

\section{Other key-recovery routes and the scope of the objections}\label{sec:mceliece-routes}

\paragraph{Weis's comparison of the routes.} Weis compares the algorithms built on the GIJS distinguisher by three features: which kernel of $E$ they read, what they read from it, and how they resolve the Frobenius ambiguity, that is, the choice among the $m$ conjugate curves through the binary points \cite[Section 8 and Table 6]{Wei26}. His comparison refers to the versions of the other papers available in September, in particular Vedenev's working draft of September 6 and Saarinen's costing of September 15. The key recovery of GIJS reads the canonical relation in the column kernel and fixes the branch by feasibility tests at a fourth held-out column; because it does not separate the canonical relation from the local ones, it needs many large solves over $\F_q$, and its cost is given as $2^{130}$--$2^{149}$ for the five sets in the revised GIJS paper (\cref{sec:mceliece-wave}). Saarinen's itemised analysis of the same algorithm gives $2^{126.8}$--$2^{145.2}$ bit operations with working storage of $2^{51.3}$--$2^{59.1}$ bits (\cref{tab:saarinen-cost}) \cite{Saa26cost}. Vedenev's two attacks of that version read the row kernel, as the gradient attack does. His minor-code attack uses tangent spaces at every column and, under a completeness conjecture, isolates a GRS component from an auxiliary code built from $2\times2$ minors, which is then attacked as in Sidelnikov--Shestakov; his jet-consistency attack uses deep derivative flags at a few columns and guesses their support elements and relative Frobenius twists. Both were run end to end on Goppa codes over $\F_4$, without figures per parameter set. According to the BSI's account, the author regarded the algorithm as applicable to Classic McEliece parameter sets, at the price of several runs of the GIJS distinguisher, whose cost is then a lower bound \cite{BSI26}. Weis notes that the minor-code attack uses the same data as his gradient attack and has the same dominant cost, the runs needed to hold out every column. Weis's own attacks cost $2^{100.8}$--$2^{110.2}$ (gradients) and $2^{94.3}$--$2^{102.4}$ (single run) in GIJS's formulas, and $2^{89.4}$--$2^{98.4}$ for the single run in his count. His summary of the trade-off is that deep flags at a few columns need few runs but do not fix the global Frobenius labelling; first-order data at every column make the labelling a linear problem but need $\lceil n_\ell/H\rceil$ runs; and the column-kernel routes avoid both costs, provided the local relations are removed explicitly. Weis also records that the single-run attack is built on the central observation of the GIJS key recovery, and that the revised paper of GIJS lists several of his improvements, including the single-run reading, as suggestions.

\paragraph{Apon's bound.} Apon's August analysis \cite{Apo26} challenges the independence of held-position equations in the earlier jet-consistency route, at the step that reconstructs the hidden curve from derivative flags at held positions. Under its nondegeneracy condition, the required count exceeds $2t+3$; for \texttt{mceliece8192128} this gives more than 259 held positions and a cost above $2^{1500}$ for that method. It is not a lower bound on all Holdout attacks: neither of Weis's attacks contains such a step \cite{Wei26}.

\paragraph{Vedenev's October revision and the scope of the objections.} Vedenev's revision of October 7 \cite{Ved26} replaces the earlier finishing steps by two one-run routes: one reconstructs tangent planes and a GRS minor code; the other obtains an equivalent support through ratios of quadratic forms. It reports recoveries of TII-173, TII-249 and TII-253 and conditional estimates of $2^{89.9}$--$2^{98.6}$ bit operations, close to Weis's single-run count. The author states that Apon's objection does not apply to these revised routes. This is a version-specific claim, not an independent validation here: rank, solver and finishing assumptions still require assessment at Classic McEliece sizes, and neither Weis's comparison nor the BSI note covers this revision. Similarly, Saarinen's critique of a conjecture in Vedenev's September 6 draft \cite{Saa26cost} must not be presented as a refutation of every later revision.

\section{Assessment}\label{sec:mceliece-verdict}

\paragraph{What Weis's paper establishes.} We draw six conclusions. First, on its status: Weis's paper is the most complete quantitative case so far that, if its heuristic assumptions hold at full size, the hidden Goppa structure, not generic decoding, is the weakest point of every Classic McEliece parameter set. In each cost model in which both have a figure, one run is cheaper than ISD, by $41$ to $59$ bits for \texttt{mceliece348864} and by $77$ to $193$ bits for the other sets; the single-run route needs one run and the gradient route between $103$ and $1443$. It is not a demonstrated attack: four heuristic assumptions, tested on instances with at most about $10^7$ unknowns, carry figures for instances with $2^{38}$ to $2^{42}$ unknowns, and the paper itself states that none of it is close to practical and that a run at full size cannot be tested. What distinguishes it from the estimates that preceded it is that it fixes one computation, counts its operations, traffic and memory consistently, proves two ingredients that had been assumptions, and states which figure rests on which assumption.

Second, the cost model is now the main quantitative uncertainty. For one and the same run, the figures of \cref{tab:weis} differ by $28$ to $30$ bits between the cheapest model (operations with zeros skipped, memory free) and the most expensive (the guide's rule with $\gamma=1/2$). For the four larger sets this spread does not change the conclusion: even under the guide's rule the run stays at least $82$ bits below the NIST level of its category, and without a memory charge it would stay far below even if Assumption 1 failed. For \texttt{mceliece348864} the margin below its level, $51$ bits without a memory charge and $26$ to $43$ bits with one \cite{Wei26}, is of the same order as the spread itself, so the size of its shortfall, though not the existence of one, depends on the model.

Third, post-quantum security levels are defined against quantum adversaries, so the comparison must include quantum attacks. Grover's algorithm applied to Prange's ISD needs about $2^{101}$, $2^{124}$, $2^{164}$, $2^{164}$ and $2^{183}$ logical gates for the five sets, ignoring error correction, memory and depths of at least $2^{71}$ to $2^{150}$ \cite{Ber10,Wei26}. For \texttt{mceliece348864} the classical run is cheaper than this without a memory charge (by $9$ bits) and, marginally, with a cube-root charge per access, but $6$ to $16$ bits more expensive with the square-root charges, so for this set the structural attack and quantum generic decoding are of comparable cost. For the four larger sets the run stays below the quantum gate count, except under the guide's rule for \texttt{mceliece460896}, where the two are level (the run is $0.7$ bits above in Weis's unrounded figures). Weis knows of no quantum speed-up for a run \cite{Wei26}; whether quantum algorithms could accelerate this kind of sparse linear algebra over $\F_2$ is open (\cref{q:quantum}). If the assumptions hold, the lowest published cost estimate for every Classic McEliece parameter set in a world with large quantum computers is therefore that of a classical structural attack, or comparable to that of the best quantum generic attack. This conditional statement, not a demonstrated break, is the precise sense in which the concrete security of Classic McEliece is under revision; the companion survey records it in this sense \cite{FV26survey}.

Fourth, on the BSI's figure. The range $2^{99}$--$2^{102}$ that the BSI quotes from \cite{Wei26} for the parameter sets of TR-02102-1 \cite{BSI26} coincides with the single-run (b) figures of \cref{tab:weis}, the single-run key recovery costed with GIJS's formulas, which charge two operations per nonzero entry and nothing for memory, for the four parameter sets other than \texttt{mceliece348864} ($2^{99.3}$--$2^{102.4}$). For the same four sets Weis's own count is lower ($2^{95.4}$--$2^{98.4}$ operations, or $2^{98.3}$--$2^{101.3}$ gates) and his best memory-charged estimate higher ($2^{114.0}$--$2^{117.2}$). The BSI's caution about memory-access costs thus points at the axis along which the published figures vary most, but no model considered by Weis brings these sets near their targeted levels.

Fifth, on practicality. The estimates are certificational. A run is a chain of $2^{30}$ to $2^{34}$ dependent iterations, each an exchange among all parts of a machine with $100$ TiB to $1.5$ PiB of memory; a mesh at $10^9$ steps per second would need between $7$ and $420$ years, and additional hardware buys little time. What has been shown is a credible conditional case that the claimed security levels fail, the concern behind the BSI's recommendation, and not an imminent practical break. But the history of 2026, from the $2^{3000}$ of support splitting to these figures within months, gives little reason to treat $2^{107}$--$2^{117}$ as an end point, and the BSI expects further improvements \cite{BSI26}.

Sixth, on repair. Because the cost of a run is driven by $m$, binary Goppa codes with $n<2^m$ and larger $m$ might restore the margins at public keys $2.5$ to $6.8$ times larger \cite{Wei26}; but such parameter sets extrapolate Assumptions 1 and 4 and ignore other attacks, and they would form a new design whose conservatism would have to be re-established. For deployment this supports the ordering that the BSI recommends: HQC, which is not exposed to these Goppa-specific attacks but needs its own security analysis, as the code-based component to analyse first, and Classic McEliece, if used at all, as an additional component pending re-evaluation.

\paragraph{How the evidence fits together.} \Cref{tab:mceliece-evidence} separates the questions that are easily conflated. It also shows why the absence of a practical attack and the BSI's caution are compatible.

\begin{table}[htbp]
\centering\small
\renewcommand{\arraystretch}{1.15}
\begin{tabularx}{\textwidth}{@{}L{3.15cm}Y Y@{}}
\toprule
\textbf{Evidence} & \textbf{What it supports} & \textbf{What remains separate}\\
\midrule
GIJS distinguisher \cite{GIJS26} & Recognition of Goppa public keys in the stated asymptotic regime. & Efficient decryption and equivalent-key recovery need additional arguments.\\
HOVER and TII recoveries \cite{Saa26a,Saa26b,Wei26,Ved26} & Demonstrated recovery of particular small public challenge keys. & Success and resources for a Classic McEliece parameter distribution.\\
Full-size cost analyses \cite{Saa26cost,Wei26,Ved26} & Conditional arithmetic estimates for specified algorithms. & Structural success, repetitions, memory costs and implementation validation.\\
Fourier-sampling limitations \cite{DMR11} & Limits on specified quantum measurements in a code-equivalence formulation. & Security against classical structure recovery or arbitrary quantum algorithms.\\
BSI guidance \cite{BSI26} & A dated deployment recommendation responding to the developing evidence. & A practical key-recovery demonstration or a general security theorem.\\
\bottomrule
\end{tabularx}
\caption{Evidence categories for Classic McEliece, as checked on October 11, 2026. No row by itself settles the other questions.}
\label{tab:mceliece-evidence}
\end{table}

\section{Standardisation, deployment guidance and consequences}\label{sec:deployment}

\paragraph{Standardisation and the BSI assessment.} NIST did not select Classic McEliece at the end of its fourth round \cite{NISTIR8545}. ISO/IEC standardised selected parameter sets in June 2026, in ISO/IEC 18033-2:2006/Amd 2:2026 \cite{ISO18033Amd2,ClassicMcElieceISO}; these exclude the smallest submitted set, \texttt{mceliece348864}. Standardisation is distinct from an ongoing cryptanalytic assessment. The BSI note of 29 September 2026 records this standardisation and the BSI's recommendation of Classic McEliece since 2020 \cite{BSI26}. Its assessment is that the published results do not yet enable a practical attack on the parameter sets recommended in TR-02102-1, that the estimates and cost models of the most recent, unreviewed works must be checked carefully, for example with regard to memory-access costs, and that a significant improvement in structural attacks has nevertheless occurred, with further improvements ``conceivable and to be expected''. Its recommendations are that Classic McEliece should no longer be used in new developments; that key agreement should use FrodoKEM, ML-KEM or, once its standard is published, HQC; that the current recommendation for Classic McEliece be limited to hybrid use with a classical scheme, combined so that both must be broken to obtain the key; and that the recommendation will be revised in the next version of TR-02102-1, expected in early 2027. The note states that FrodoKEM, ML-KEM and HQC are not affected by these results \cite{BSI26}. Its numerical estimates predate some of the preprint revisions above and should not be mixed with them. A classical component can preserve a classical security guarantee if the combiner has the required property; it alone cannot provide a post-quantum guarantee against Shor's algorithm.

\paragraph{Quasi-cyclic codes: HQC and BIKE.} HQC works in $\F_2[x]/(x^n-1)$ with $n$ prime. The secret key is a pair of sparse polynomials $(x,y)$ and the public key is $(h,s=x+hy)$ for uniform $h$; to encrypt $\mu$, one samples sparse $r_1,r_2,e$ and sends $u=r_1+hr_2$ and $v=\mu G+sr_2+e$, where $G$ generates a \emph{public} decodable code; the receiver computes $v-uy=\mu G+(xr_2-r_1y+e)$, whose error term has low weight, and decodes \cite{HQC}. Since the decoding code is public, security rests on the hardness of decoding random quasi-cyclic codes rather than on hiding a structured code. NIST selected HQC for standardisation in March 2025 \cite{NISTIR8545}, with a draft standard (FIPS 207) in preparation. BIKE uses quasi-cyclic moderate-density parity-check codes: the secret is a pair of sparse polynomials $(h_0,h_1)$, the public key is $h=h_1h_0^{-1}$, and decryption uses bit-flipping decoding, whose failure rate must be negligible for CCA security \cite{BIKE}; it was not selected \cite{NISTIR8545}. Quasi-cyclic codes are the code-based analogue of ideal lattices, and structured decoding requires its own analysis (\cref{q:structured}).

\paragraph{Consequences.} The attacks surveyed here are classical algebra, the solving of algebraic models, large sparse linear systems over relation spaces, algebra computations and polynomial factorisation, with no hidden-shift structure. They leave untouched the position of Classic McEliece with respect to quantum hidden-structure attacks (\cref{sec:bg-quantum}). What changes is the evidence on its concrete security: the lowest published estimates are no longer those of generic decoding with a Grover speed-up but those of classical structural attacks whose concrete cost is under revision, conditionally cheaper than quantum generic decoding for the four larger parameter sets in essentially every cost model considered, and comparable to it for \texttt{mceliece348864} \cite{Wei26}. This is neither an endorsement of new deployments nor a claim of a practical full-size break. The lower conditional estimates warrant scrutiny of the structural assumptions and of realistic memory costs; the challenge recoveries show that some recovery mechanisms work, but not that their success or resources extrapolate to the standard sets. Three consequences follow. First, among standardised and standard-track KEMs, HQC is the code-based KEM that, according to the BSI, is not affected by these attacks \cite{BSI26}; its decoding code is public and its security rests on decoding random quasi-cyclic codes, so it needs its own security analysis, and its lack of exposure to these particular Goppa attacks is not a blanket security guarantee. Second, hybrid designs that seek code-based diversity would take HQC as the code-based component to analyse first, with Classic McEliece used, if at all, as an additional component pending re-evaluation, in line with the BSI's advice; any such combination still needs a combiner with the required security property \cite{GHP18,BBFGS19}. Third, the episode illustrates again that weaknesses tend to appear where algebraic structure is hidden, in ideal lattices \cite{CDPR16}, in overstretched NTRU \cite{ABD16}, and now in the Goppa structure of McEliece, while HQC, whose decoding code is public, is not affected by these attacks.

\section{Open questions and conclusions}\label{sec:open}

The following questions would settle what the evidence of \cref{sec:mceliece-wave,sec:mceliece-saarinen,sec:mceliece-weis,sec:mceliece-routes} leaves open.

\begin{question}[Success and cost at full size]\label{q:fullsize}
Do the Holdout/Hermite key-recovery methods \cite{GIJS26,Saa26cost,Saa26b,Ved26,Wei26} succeed on the Classic McEliece parameter distributions, with what structural success probabilities and end-to-end costs, including repetitions and realistic memory access? In particular, do the heuristic assumptions behind the lowest estimates, on the dimension of the row kernel, the span of the gradients, the structure of the column kernel and the behaviour of block Lanczos over $\F_2$ \cite{Wei26}, hold at $2^{38}$ to $2^{42}$ unknowns?
\end{question}

\begin{question}[From challenges to parameter sets]\label{q:extrapolation}
Can the reported small-instance recoveries be extended without assuming the full-size rank and locator-uniqueness properties? For the two-anchor method, what is the probability over key generation that two complete conditioned kernels expose the required field geometry, and that enough compatible locators can be obtained for economical key completion \cite{Saa26b}?
\end{question}

\begin{question}[Memory and parallelism]\label{q:memory}
What does a run cost under realistic models of memory access and communication? Weis observes that its long-distance traffic does not depend on the block width, so that, unlike exhaustive key search or ISD with small lists, a run does not keep its cost when it is spread over more hardware \cite{Wei26}. Is there an organisation of the computation that avoids this?
\end{question}

\begin{question}[Quantum acceleration]\label{q:quantum}
Does a quantum computer accelerate the sparse linear algebra over $\F_2$ that dominates these attacks? Weis knows of no quantum speed-up for a run \cite{Wei26}. For \texttt{mceliece348864}, where the structural run and Grover-accelerated ISD are of comparable cost (\cref{sec:mceliece-verdict}), the answer would decide which attack is cheaper in a world with large quantum computers.
\end{question}

\begin{question}[Repair]\label{q:repair}
Is there a regime of binary Goppa codes, for example with $n<2^m$ and larger $m$, or another family of codes with an efficient decoder that resists all known structural attacks with public keys of comparable size? Weis's example sets with $n<2^m$ extrapolate his assumptions far beyond the tests and ignore other attacks at large $m$; he presents them as illustrations, not proposals \cite{Wei26}.
\end{question}

\begin{question}[Structured decoding]\label{q:structured}
Is there a hidden-structure attack on quasi-cyclic syndrome decoding, the problem behind HQC and BIKE, analogous to the structural attacks on hidden Goppa codes or to the quantum attacks on principal ideals in cyclotomic fields \cite{CDPR16,BS16}?
\end{question}

\paragraph{Conclusions.} Within a few months of 2026, structural cryptanalysis moved the lowest published estimates for key recovery on Classic McEliece from more than $2^{3000}$ to conditional per-run figures near $2^{100}$, and it turned public challenges with brute-force labels up to $254$ into solved instances. The figures are conditional: they rest on heuristic assumptions tested on instances tens of thousands to hundreds of thousands of times smaller than the target, and they vary by about thirty bits with the cost model. Within that spread, the four larger parameter sets fall far below their targeted levels in every model considered, and \texttt{mceliece348864} falls below its level by an amount comparable to the spread. No full-size key recovery has been demonstrated, and Weis notes that no machine of the kind a run would require exists. The BSI's response, no use in new developments, hybrid use only, and a re-evaluation expected in early 2027, matches this evidence. The open questions above, and the next revisions of the preprints discussed here, will decide whether the remaining gap between conditional estimates and a demonstrated attack closes.
\bibliographystyle{splncs04}
\bibliography{Bibliography}

\end{document}